\documentclass[aps,prl,reprint,groupedaddress]{revtex4-1}
\usepackage[pdftex]{graphicx}
\usepackage{amsmath}

\begin{document}

%Title of paper
\title{Direct observation of the coherent nuclear response after 
the absorption of a photon}
%Real time observation of the Conical Intersection structure for the primary event in Vision}

\author{M. Liebel}
\author{C. Schnedermann}
\author{G. Bassolino}
\author{G. Taylor}
\altaffiliation{Department of Biochemistry, Biomembrane Structure Unit, University of Oxford, Oxford, OX1 3QU, United Kingdom.}
\author{A. Watts}
\altaffiliation{Department of Biochemistry, Biomembrane Structure Unit, University of Oxford, Oxford, OX1 3QU, United Kingdom.}
\author{P. Kukura}
\email{philipp.kukura@chem.ox.ac.uk}
\affiliation{Physical and Theoretical Chemistry Laboratory, South Parks Road, Oxford OX1 3QZ, UK}

\begin{abstract}
How molecules convert light energy to perform a specific transformation is a fundamental question in photophysics. Ultrafast spectroscopy reveals the kinetics associated with electronic energy flow, but little is known about how absorbed photon energy drives nuclear or electronic motion. Here, we used ultrabroadband transient absorption spectroscopy to monitor coherent vibrational energy flow after photoexcitation of the retinal chromophore. In the proton pump bacteriorhodopsin we observed coherent activation of hydrogen wagging and backbone torsional modes that were replaced by unreactive coordinates in the solution  environment, concomitant with a deactivation of the reactive relaxation pathway. 
\end{abstract}

%\maketitle must follow title, authors, abstract, \pacs, and \keywords
\maketitle

%\section{Introduction}
The efficiency of light-induced processes fundamentally depends on the ability of a system to convert photons into chemical, electrical or mechanical energy. Resonance Raman intensity analysis reveals the initial energy distribution and the systems evolution over the first few tens of femtoseconds after photoexcitation \cite{Warshel1977,Heller1981}, while time-resolved optical spectra provide electronic kinetics. Neither of these techniques is capable of following the evolution of incident photon energy from the initially populated Franck-Condon (FC) region toward reactive rather than dissipative channels. It is this energy transfer, however, that is critical in determining the efficiency and outcome of a photo-induced process \cite{Takeuchi2008} and has been suggested to be subject to quantum coherent effects even for large protein assemblies at ambient condition \cite{Qing1994,Engel2007,Collini2010}.

The variation in reaction yields and speeds for retinal protonated Schiff base (RPSB) photoisomerization in different environments highlights the importance of the competition between active and passive relaxation pathways. In the proton pump bacteriorhodopsin (bR), excited state decay following photoexcitation is faster (0.5 vs 4 ps) \cite{Mathies1988,Kandori1993}, results in a higher isomerization yield (0.64 vs 0.16) \cite{Tittor1990,Freedman1986} and produces exclusively the 13-\textit{cis} rather than the 11-\textit{cis} isomer formed in solution. Nevertheless, the resonance Raman spectra of both RPSBs are strikingly similar \cite{Smith1987} implying an essentially identical starting geometry and initial structural evolution. Raman active totally symmetric stretching coordinates, however, cannot contribute to the formation of a conical intersection between states of opposite symmetry and thus to efficient electronic relaxation \cite{Levine2007}.

\begin{figure}[ht]
\includegraphics{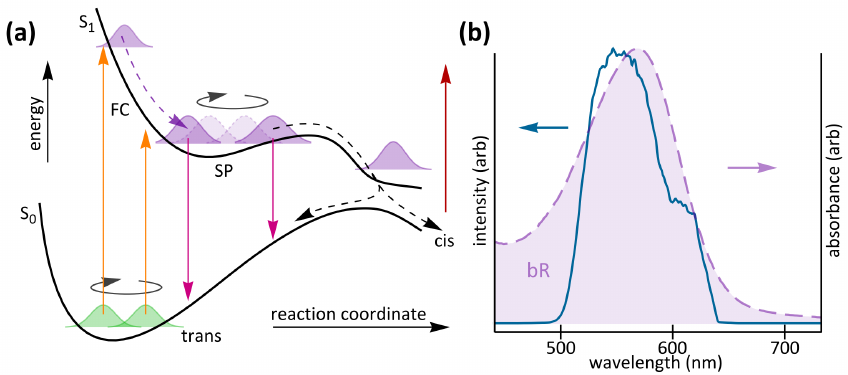}
\caption{\label{surface} Schematic potential energy diagram for retinal isomerization and excitation conditions. (a) Potential energy of the ground (S$_0$) and excited (S$_1$) electronic states of RPSB along the isomerization coordinate including nuclear wavepacket dynamics. FC: Franck-Condon region; SP: stationary point. (b) Overlay of the 9.6 fs excitation pulse (solid) with the bacteriorhodopsin absorption spectrum (dashed).}
\end{figure}

The reactivities observed for RPSB in solution compared to the protein environment must therefore result from the activation of different nuclear motions by energy transfer from the initially populated coordinates. To reveal the energy flow, we created and followed vibrational coherence with specific sensitivity to the region near the vibronic surface crossing using impulsive vibrational spectroscopy \cite{Fragnito1989}. Illustrating this experimental approach with RPSB, absorption of an ultrashort visible photon generates vibrational coherence in all nuclear degrees of freedom coupled to the optical transition in both ground (S$_0$) and excited (S$_1$) electronic states (FIG. \ref{surface}) \cite{Pollard1992}. The impulsively created S$_0$ nuclear wavepacket oscillates about its equilibrium position whereas on S$_1$  it rapidly moves out of the FC region toward a new stationary point (SP) \cite{Garavelli1997}. At this local minimum, the backbone structure has evolved little along the isomerization coordinate and experiences a barrier toward excited state decay \cite{Hasson1996,Ruhman2002}. Previous attempts to reveal reactive nuclear wavepacket evolution \cite{Kobayashi2001} have struggled either with interfering ground state signatures \cite{Dexheimer1992,Kraack2011} or insufficient time-resolution revealing only low-frequency coherences \cite{Wand2012}. As a result, only limited insight into the reactive dynamics from the perspective of a few degrees of freedom was available. 

Here, we combined high time resolution transient absorption spectroscopy with ultrabroad (500 - 900 nm) detection bandwidth, spectroscopic sensitivity ($<$10 $\mu$OD) and electronic population control to conclusively reveal excited state nuclear wavepacket dynamics over the full spectral window of interest ($<$2000 $\mathrm{cm}^{-1}$).

\begin{figure}[ht]
\includegraphics{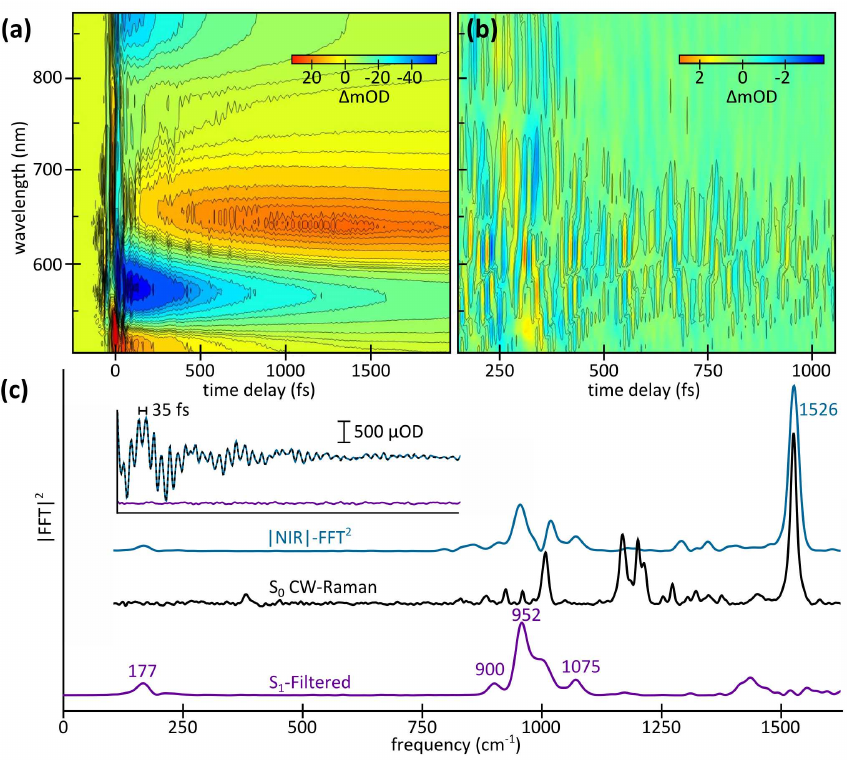}
\caption{\label{maps} Reactive nuclear wavepacket dynamics during retinal isomerization in bacteriorhodopsin (bR). (a) Chirp corrected \cite{Kovalenko1999} differential absorbance map of bR after resonant excitation by a 9.6 fs pulse (FIG. \ref{surface}b). The sample was flown through a sample cell at OD = 0.45/500 $\mu$m. Pump and probe intensities were adjusted to be 200 and 2.5 nJ at a laser repetition rate of 2 kHz with 100 and 60 $\mu$m beam diameters, respectively. (b) Residual vibrational coherence map after global subtraction of the exponential electronic kinetics \cite{Liebel2013}. The contour lines shown were limited to the 800-1200 cm$^{-1}$ frequency window for clarity. (c) Fourier power spectrum of the coherence after application of a Kaiser window before and after removal of long-lived ($>$0.45 ps dephasing time) coherences and comparison to S$_0$ Raman spectrum. The peak intensities were corrected for the limited time-resolution. Inset: wavelength averaged (760-840 nm) coherent activity of b, fit to 24 exponentially decaying oscillations.}
\end{figure}

The differential absorbance map of RPSB in bR after optical excitation with a $<$10 fs pulse is shown in FIG. 
\ref{maps}a. Early time-delays ($<$100 fs) are dominated by the coherent artifact caused by the interaction of pump and probe pulses with the sample \cite{Dobryakov2005}, rapidly followed by ground state bleach (570 nm) and stimulated emission bands ($>$750 nm). Both signatures decay with the characteristic excited state lifetime of 0.5 ps concomitant with a growth of the photoproduct absorption band at 640 nm. Given the high time-resolution ($<$20 fs) achieved in this work, the broad and slowly varying spectral and temporal electronic features are complemented by rapid oscillations caused by the impulsively excited nuclear wavepackets (see FIG. \ref{surface}a). 

Subtraction of the exponential electronic kinetics reveals the pure vibrational coherence, including two phase jumps, one at slightly lower energy than the S$_0$ absorption maximum (580 nm) and another at the border between ground and excited state signatures (750 nm) (FIG. \ref{maps}b). The near-infrared (NIR) coherence is strongly damped compared to the one observed in the S$_0$ absorption window ($<$650 nm), in agreement with the short  S$_1$ lifetime (0.5 ps). These observations suggest that the visible region is dominated by nuclear coherence on S$_0$, while the NIR mainly exhibits signatures caused by vibrational wavepackets evolving on S$_1$.

To conclusively remove any residual S$_0$ coherences in the NIR, we determined the frequencies and dephasing times of all contributing oscillations using a combination of linear prediction singular value decomposition \cite{Johnson1996} and nonlinear optimization. A sum of 24 exponentially decaying oscillations was sufficient to describe the observed coherence decay to within the measurement noise (FIG. \ref{maps}c inset) and revealed the presence of some coherences with dephasing times exceeding 0.45 ps, all of which showed frequencies matching known S$_0$ bands, such as the C=C band at 1526 cm$^{-1}$. 

After removal of these S$_0$ signatures we are left with a spectrum representative of coherence activity on S$_1$ only exhibiting a completely altered intensity and frequency distribution compared to the S$_0$ frequency-domain Raman spectrum of bR (FIG. \ref{maps}c). The hydrogen wag at 952 cm$^{-1}$ dominates the spectrum, with lesser contributions from additional wagging (900 cm$^{-1}$), Schiff base (1075 cm$^{-1}$) and backbone torsional (177 cm$^{-1}$) coordinates. Recent coherent infrared emission spectra on bR agree with our observations of strong activity in the 750-1150 cm$^{-1}$ region, but an unambiguous assignment of the observed features was hampered by numerous additional modes of unknown origin \cite{Groma2011}. 

An important consequence of our detection methodology is that vibrational motion which strongly modifies the S$_0$-S$_1$ energy gap will also generate the most prominent oscillatory features in the transient electronic spectra (FIG. \ref{surface}a). Given that the critical step during the isomerization reaction is the passage from S$_1$ to S$_0$ through a conical intersection, the most active modes in our experiment are likely to be the strongest contributors to excited state decay and thus the photoisomerization. Despite the complexity of the system with more than 300 nuclear degrees of freedom, the S$_1$ vibrational coherence of RPSB in bR is dominated by the hydrogen wag at 952 cm$^{-1}$, a motion essential to C=C isomerizations \cite{Kukura2005} and capable of coupling the ground and excited electronic states potential energy surfaces (PES). The results presented in FIG. \ref{maps}c thus suggest a correspondence between the ability of a systems to focus energy into a few critical coordinates and the photochemical outcome of excited state decay.

\begin{figure}[ht]
\includegraphics{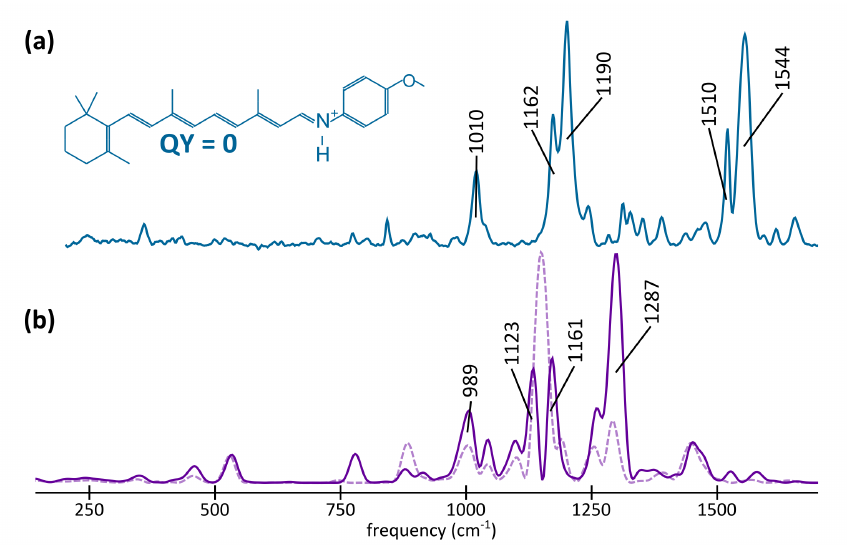}
\caption{ \label{pma} Comparison of ground and excited state Raman activity in p-methoxyaniline RPSB (pma-RPSB). (a) Molecular structure of pma-RPSB and ground state frequency-domain Raman spectrum recorded in acetonitrile. (b) Fourier power spectrum of photon-induced excited state vibrational coherence of pma-RPSB (solid) and fully deuterated pma-RPSB (dashed) averaged over the 760 - 850 nm spectral window measured analogous to RPSB in bR using a resonant 6.8 fs pulse. For pma-RPSB, no removal of residual ground state coherences as for RPSB in bR was possible, but the dramatically different appearance of the ground and excited state spectra suggests minor contributions from residual S$_0$ coherences.}
\end{figure}

Although such an interpretation is appealing, it does not prove a correlation between the observed nuclear coherence, the shape of the PES and photochemical reactivity. By synthesizing a series of RPSBs with different Schiff bases, we found that using p-methoxyaniline (pma) as a base completely eliminated all isomerizing channels (FIG. \ref{pma}a). The close resemblance of the S$_0$ Raman spectrum of pma-RPSB to that of RPSB in bR (see FIG. \ref{maps}c) suggests similar ground state geometries, excited electronic state character and initial nuclear evolution for RPSB in bR and pma-RPSB in solution. In contrast to RPSB in bR, however, the S$_1$ coherence shows negligible activity in the hydrogen wagging ($<$1000 cm$^{-1}$) and torsional ($<$300 cm$^{-1}$) regions of the spectrum and is instead dominated by stretching coordinates with significant contributions likely originating from the aromatic Schiff base moiety (1000-1300 cm$^{-1}$), as verified by comparison with fully deuterated pma (FIG. \ref{pma}b).

In contrast to most current ultrafast techniques, the vibrational coherence presented in this work was generated by the photon that triggers the light induced process. Our results thus do not simply report on the structural evolution of the molecule of interest, but instead on the coherent vibrational energy flow along the S$_1$ PES out of the FC region during the isomerization reaction. Based on the S$_0$ Raman spectra, we would have expected strong coherent activity in backbone stretches for RPSB in bR and in solution, but observed exactly the opposite experimentally. In fact, given that S$_1$ vibrational coherence is generated mainly in FC active modes, we would expect the power spectra of S$_1$ coherent activity to be as similar as the respective S$_0$ Raman spectra. 

The dramatically different appearance of the S$_0$ and S$_1$ spectra cannot be solely attributed to vibrational structural changes caused by different bonding properties upon electronic excitation. The apparent lack of backbone stretching activity in bR is a \textit{priori} surprising, given the dominant presence of C=C stretching modes in the Raman spectra of polyenic systems. Such behaviour, however, has been previously observed in the photoisomerization of stilbene \cite{Dobryakov2012}. Here, totally symmetric coordinates remain dominant in the Raman spectrum of the excited electronic state for the \textit{trans}-isomer but disappear almost completely for the more reactive \textit{cis}-isomer likely due to significant symmetry changes caused by molecular distortion. The key question thus relates to the mechanism by which vibrational coherence can appear in initially silent and disappear from initially excited degrees of freedom. The latter can be explained by a change in equilibrium displacement between S$_0$ and S$_1$ as the structure of the molecule evolves on the excited electronic state. The former, however, must be related to the shape of the PES away from the FC region \cite{Takeuchi2008}. 

\begin{figure}[ht]
\includegraphics{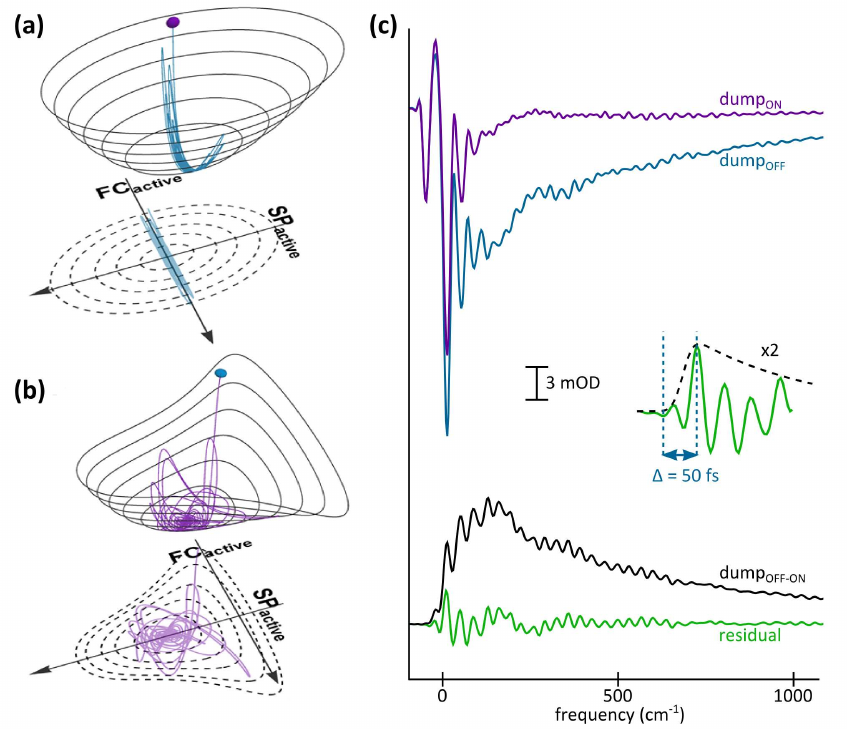}
\caption{\label{coupling} Anharmonic coupling mediates coherent activation of hydrogen wagging on the $<$50 fs timescale. (a) Potential energy surface and classical trajectory for two uncoupled nuclear coordinates, with only one coordinate being significantly displaced with respect to the ground electronic state. (b) PES and dynamic evolution for identical initial displacements as in a, but with the addition of a strong anharmonic coupling term. (c) Recording the differential absorption in the 760-840 nm spectral region in the presence and absence of a ps dump pulse resonant only with the S$_0$$\leftarrow$S$_1$ transition removes the coherent artifact around zero time delay and reveals the rapid rise of the coherent vibrational activity. Inset: Magnification of the rising coherence amplitude.}
\end{figure} 

Consider a simplified, two-dimensional excited state PES consisting of two uncoupled nuclear coordinates, only one of which is displaced from its equilibrium position in the FC region. For a $\pi^*\leftarrow\pi$ excitation, two such coordinates may be the symmetric C=C stretch and an asymmetric hydrogen wag. Upon population of S$_1$, the system oscillates along the displaced stretching coordinate about its new  equilibrium bond length but without gaining any momentum along the initially silent asymmetric wag (FIG. \ref{coupling}a). By introducing a coupling term between the two modes this behavior changes drastically. The system now gains momentum along the asymmetric coordinate after the first tens of fs, even if the initial displacement was close to zero (FIG. \ref{coupling}b). Such a shift, however, does not necessarily induce significant changes in the resonance Raman spectra, because only the earliest dynamics are sampled \cite{Heller1981}.

To quantify the timescale for the coherent vibrational coupling process, we used electronic population control to eliminate the otherwise dominant coherent artifact around time-zero, thus revealing the full temporal evolution of the vibrational coherence down to the earliest time-delays. Using a long (1 ps) dump pulse centered at 880 nm, we selectively removed S$_1$ population without impulsively generating additional nuclear wavepackets or affecting the S$_0$ population. By subtracting the transient dynamics in the absence and presence of the dump pulse we removed the coherent artifact and revealed the rapid rise of the vibrational coherence on the sub 50 fs timescale (FIG. \ref{coupling}c). Interestingly, this growth is comparable to the recently reported rise time of the SE signal and matches the kinetics of tentatively assigned excited state signatures at wavelengths $>$1 $\mu$m \cite{Hasson1996,Wand2012}. The major changes responsible for the observed vibrational coherence transfer does take place during the first tens of femtoseconds after photoexcitation.

In conclusion, we used ultabroadband transient absorption spectroscopy with high temporal resolution to reveal the coherent nuclear response of a molecule after absorption of a photon. In contrast to the expectation that the photon-induced structural evolution largely takes place along FC active coordinates, we observed efficient and coherent activation of initially silent vibrational modes caused by PES-induced anharmonic coupling. We further demonstrated that the nature of the coherent evolution differs dramatically with the environment of the molecule and is strongly correlated with the efficiency and outcome of the light induced process as evidenced by the comparison of RPSB in bR and pma-RPSB in solution.  Interestingly, the coherent evolution takes place on the $<$50 fs timescale, fast enough to allow for quantum coherent effects to remain active and decisive for the eventual outcome of the process \cite{Prokhorenko2006}, irrespective of the photoexcitation mechanism. Our results emphasize the importance of ultrafast energy flow in determining the outcome of light-induced processes and demonstrate the power of vibrational spectroscopy in the time-domain for studying the structural and dynamic origins of molecular relaxation phenomena.

\bibliography{bR_PRL_v1.0}
\end{document}